\documentclass[prl,twocolumn,aps,showpacs]{revtex4}
\usepackage{amssymb}
\usepackage{graphicx}
\usepackage{amsmath}
\usepackage{amsfonts}
\begin{document}  
\author{Eva Pavarini}
\author{Erik Koch}
\affiliation{Institut f\"ur Festk\"orperforschung and
             Institute for Advanced Simulation,
             Forschungzentrum J\"ulich,  52425 J\"ulich, Germany}
\date{\today}
\title{Origin of Jahn-Teller distortion and orbital-order in LaMnO$_3$}
\begin{abstract}
The origin of the cooperative Jahn-Teller distortion and orbital-order in LaMnO$_3$ is central to the physics of the manganites. The question is complicated by the simultaneous presence of tetragonal and GdFeO$_3$-type distortions and the strong Hund's rule coupling between $e_g$ and $t_{2g}$ electrons. To clarify the situation we calculate the transition temperature for the Kugel-Khomskii superexchange mechanism by using the local density approximation+dynamical mean-field method, and disentangle the effects of super-exchange from those of lattice distortions. We find that super-exchange alone would yield $T_{\rm KK}\!\sim\!650$~K. The tetragonal and GdFeO$_3$-type distortions, however, reduce $T_{\rm KK}$ to $\sim\! 550$~K. Thus electron-phonon coupling is essential to explain the persistence of local Jahn-Teller distortions to  $\gtrsim 1150$~K and to reproduce the occupied orbital deduced from neutron scattering.
\end{abstract}
\maketitle

The insulating perovskite LaMnO$_3$ is the parent compound of the colossal magneto-resistance manganites \cite{science} and it is considered a text-book example of a cooperative Jahn-Teller (JT) orbitally-ordered material \cite{book}. Two distinct mechanism have been proposed to explain the cooperative distortion: many-body Kugel-Khomskii (KK) super-exchange (SE) \cite{kugel} and one-electron electron-phonon (EP) coupling \cite{JT}. Determining the relative strength of these mechanisms will provide a measure of the importance of strong correlation effects for the orbital physics in the manganites. Unfortunately the situation is complicated by the simultaneous presence of tetragonal and GdFeO$_3$-type distortions as well as a strong Hund's rule coupling between the Mn $e_g$ and $t_{2g}$ electrons. 

\begin{figure}
\centerline{\includegraphics [width=0.35\textwidth]{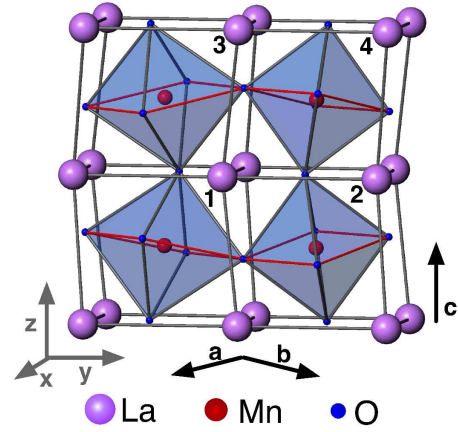}}
\caption{\label{str}
(Color online) Structure of LaMnO$_3$ at 300~K \cite{rodriguez98}. 
The conventional cell is orthorhombic with axes {\bf a}, {\bf b} and {\bf c}, and contains 4 formula units. The pseudo-cubic axes (left corner) are defined via ${\bf a}=({\bf {x}}-{\bf {y}}) (1+\alpha)$, ${\bf b}=( {\bf x}+{\bf y})(1+\beta)$, and ${\bf c}=2{\bf z} (1+\gamma)$, with $\alpha,\beta,\gamma$ small numbers. 
For sites 1 and 3 the long (short) bond $l$ ($s$) is $\sim$ along ${\bf y}$ (${\bf x}$), vice versa for sites 2 and 4 ($d$-type pattern). All Mn sites are equivalent. The symmetries that transform them into a site of type 1 are: $x\leftrightarrow y$ (site 2) $z \to -z$ (site 3), $x\leftrightarrow y$, $z\to -z$ (site 4).
}
\end{figure}

In LaMnO$_3$ the Mn$^{3+}$ ions are in a $t_{2g}^3\,e_g^1$ configuration. Due to strong Hund's rule coupling the spin of the $e_g$ electron is parallel to the spin of the $t_{2g}$ electrons on the same site. Above $T_N=140$~K the spins on neighboring sites are disordered \cite{TN}. The crystal structure is orthorhombic (Fig. \ref{str}). It can be understood by starting from an ideal cubic perovskite structure with axes {\bf x}, {\bf y} and {\bf z}: First, a tetragonal distortion reduces the Mn-O bond along {\bf z} by 2\%. The La-O and La-Mn covalencies induce a GdFeO$_3$-type distortion \cite{evad1,hero} resulting in an orthorhombic lattice with axes {\bf a}, {\bf b}, and {\bf c}, with the oxygen-octahedra tilted about {\bf b} and rotated around {\bf c} in alternating directions. Finally, the octahedra distort, with long ($l$) and short ($s$) bonds alternating along ${\bf x}$ and ${\bf y}$, and repeating along ${\bf z}$  \cite{rodriguez98,volume,dynamic,quadrup}. This is measured by $\delta_{JT}=(l-s)/((l+s)/2)$. The degeneracy of the $e_g$ orbitals is lifted and the occupied orbital, $|\theta\rangle= \cos \frac{\theta}{2} |3z^2-1\rangle + \sin \frac{\theta}{2}|x^2-y^2\rangle$, is $\sim |3l^2-1\rangle$, i.e. it
points in the direction of the long axis. Thus orbital-order (OO) is $d$-type with the sign of $\theta$ alternating along 
${\bf x}$ and ${\bf y}$ 
and repeating along ${\bf z}$.
At 300~K the JT distortion is substantial, $\delta_{JT}=11\%$, and $\theta\sim 108^o$ was estimated from neutron scattering data \cite{rodriguez98}. 
Above $T_{OO}\!\sim\!750$~K a strong reduction to $\delta_{JT}=2.4\%$ was reported \cite{rodriguez98,rxr}, accompanied by a change in $\theta$ to $\sim 90^o$ \cite{rodriguez98}.   
Recently this was, however, identified as an order-to-disorder transition \cite{dynamic,quadrup}: Due to orientational disorder, the crystal appears cubic on average, while, within nano-clusters, the MnO$_6$ 
octahedra remain fully JT distorted up to $T_{JT}\gtrsim$1150~K~\cite{quadrup}.

Model calculations based on super-exchange alone can account for $d$-type order, but yield, for the classical ground state, $\theta\sim90^o$ \cite{thetaSE}. Models of electron-phonon coupling in simple cubic perovskites instead give $\sim 120^o$ \cite{JT}. To explain the observed $\sim 108^o$, one might thus conclude that both mechanisms are of similar importance \cite{kugel}. Such models are lacking, however, a realistic description of the crystal and the calculated $\theta$ is sensitive to the choice of parameters \cite{angle,JT}.
LDA+$U$ calculations yield $\theta=109^o$ and show that Coulomb repulsion is fundamental to stabilize the Jahn-Teller distortions in the 
ground state \cite{Ku}. This might be taken as evidence that Kugel-Khomskii super-exchange is the dominant mechanism, and electron-phonon coupling, enhanced by electron localization \cite{Ku,tobepub}, merely helps. 
On the other hand, recent semiclassical many-body calculations for model cubic perovskites indicate that electron-phonon coupling is essential to explain orbital ordering above 300~K \cite{millis08}. 

While it is not obvious how well LDA+$U$ or semiclassical approaches capture the many-body nature of the KK super-exchange, it seems clear that the inclusion of the real crystal structure is crucial \cite{kugel,added,evad1,hotta}. The tetragonal and GdFeO$_3$-type distortions result in a sizable narrowing of the $e_g$ bands \cite{evad1,hero,millislda}, likely changing the relative strength of super-exchange and electron-phonon coupling. Since, in the presence of a crystal-field, Coulomb repulsion suppresses orbital fluctuations \cite{evad1,kcuf3}, they may even compete with SE and EP coupling. To identify the driving mechanism for orbital-order in LaMnO$_3$, it is thus mandatory to account for both the realistic electronic structure and many-body effects. To understand the mechanism one has to disentangle the contribution of KK super-exchange from that of the JT or  the GdFeO$_3$-type and tetragonal distortions.

In this Letter, we do this by calculating directly the Kugel-Khomskii super-exchange transition temperature, $T_{\rm KK}$, with and without tetragonal and GdFeO$_3$-type distortions. We adopt the method used successfully for KCuF$_3$~\cite{kcuf3}, based on  local-density approximation + dynamical mean-field theory (LDA+DMFT) \cite{lda+dmft}. 

First, we calculate the  electronic structure {\it ab-initio} using the $N^{th}$-order muffin-tin orbital method. 
Since the Hund's rule energy gain is larger that the $e_g$-$t_{2g}$ crystal-field splitting, 
the $t_{2g}$ bands are $\frac{1}{2}$- and the $e_g$ bands $\frac{1}{4}$-filled; 
the three $t_{2g}$ electrons behave as a  spin ${\bf S}_{t_{2g}}$ and couple to the $e_g$ electron via an effective magnetic field $h={J}S_{t_{2g}}$.  
In the paramagnetic phase ($T > T_N$=140~K) the ${t_{2g}}$ spins are spatially disordered. The minimal model to study the KK mechanism in LaMnO$_3$ is thus \cite{millis}
\begin{eqnarray}
 \nonumber
 H &=&\!\!\!\sum_{im\sigma,j m'\sigma'} \!\!\! 
   t^{i,i'}_{m,m'} u^{i,i'}_{\sigma,\sigma'} 
   c^{\dagger}_{im\sigma} c^{\phantom{\dagger}}_{i' m'\sigma'}\\
   \nonumber
   &-&h\sum_{im} (n_{im\Uparrow}-n_{im\Downarrow}) 
      +U\sum_{ im }  n_{im\Uparrow }n_{im\Downarrow} \\
   &+&\!\!\frac{1}{2}\!\sum_{im\left( \neq m'\right)\sigma\sigma'}
      \!\!\!(U-2J-J\delta_{\sigma,\sigma'}) n_{ im\sigma} n_{im'\sigma'}\;.
\label{H}
\end{eqnarray}
$c_{im\sigma}^{\dagger}$ creates an electron with spin $\sigma\!=\Uparrow,\Downarrow$ in a Wannier orbital $|m\rangle=|x^2-y^2\rangle$ or $|3z^2-1\rangle$  at site $i$, and $n_{im\sigma}=c_{im\sigma}^{\dagger}c^{\phantom{\dagger}}_{im\sigma}$. $\Uparrow$ ($\Downarrow$) indicates the $e_g$ spin parallel (antiparallel) to the ${t_{2g}}$ spins (on that site). The matrix $u$ ($u^{i,i'}_{\sigma,\sigma'}=2/3$ for $i\neq i'$, $u^{i,i}_{\sigma,\sigma'}=\delta_{\sigma,\sigma'}$) accounts for the orientational disorder  of the ${t_{2g}}$ spins \cite{millis}; $t^{i,i'}_{m,m'}$ is the LDA hopping integral from orbital $m$ on site $i$ to  orbital $m'$ on site $i'$, obtained {\it ab-initio} by down-folding the LDA bands and constructing a localized $e_g$ Wannier basis. The on-site terms $i=i'$ give the crystal-field splitting. 
$U$ and $J$ are the direct and exchange  
screened on-site Coulomb interaction \cite{fresard}. We use the theoretical estimate $J=0.75$~eV \cite{MF} and vary $U$ between $4$ and $7$~eV. 
The Hund's rule splitting was estimated {\it ab-initio} to $2JS_{t_{2g}}\sim 2.7$~eV \cite{hero}.
We solve (\ref{H}) using DMFT \cite{DMFT} or cellular DMFT (CDMFT) and a quantum Monte Carlo \cite{hirsch} solver, working with the full self-energy matrix $\Sigma_{mm'}$ in orbital space \cite{evad1}. The spectral matrix on the real-axis is obtained by analytic continuation \cite{jarrel}.  

We consider several structures: (i) the room temperature structure, R$_{11}$ with $\delta_{ JT}=11\%$, and a series  of hypothetical structures, R$_{\delta_{ JT}}$, with reduced JT distortion $\delta_{JT}$, (ii) the (average) structure found at 800~K, R$_{2.4}^{\rm 800K}$, which has a slightly larger volume than R$_{11}$ and a smaller GdFeO$_3$-type distortion, and (iii) the ideal cubic structure, I$_0$, with the same volume as R$_{11}$. For all structures we find that at each site the $e_g$  spins align to ${\bf S}_{t_{2g}}$. We calculate the orbital polarization $p$ as a function of temperature  \cite{fm} by diagonalizing the DMFT (or CMDFT) occupation matrix and taking the difference between the occupation of the most  ($|\theta\rangle$) and least ($|\theta+\pi\rangle$) filled orbital. To test the ${t_{2g}}$ spins picture we perform calculations for the 5-band ($e_g+t_{2g}$)  Hubbard model  \cite{details}. We find that it holds even at high temperatures.

\begin{figure}
 \centerline{\includegraphics[width=0.45\textwidth]{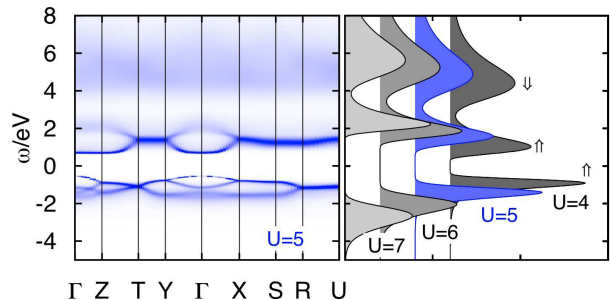}}
 \caption{\label{dmftdos} 
 (Color online) Right: LDA+DMFT spectral function for the room temperature structure R$_{11}$ for different $U$. $\Uparrow$ ($\Downarrow$) indicates states with e$_g$ spins parallel (antiparallel) to ${\bf S}_{t_{2g}}$. Left: {\bf k}-resolved spectral function for $U=5$~eV.} 
\end{figure}

\begin{figure}
 \center
 \rotatebox{270}{\includegraphics[width=1.9in]{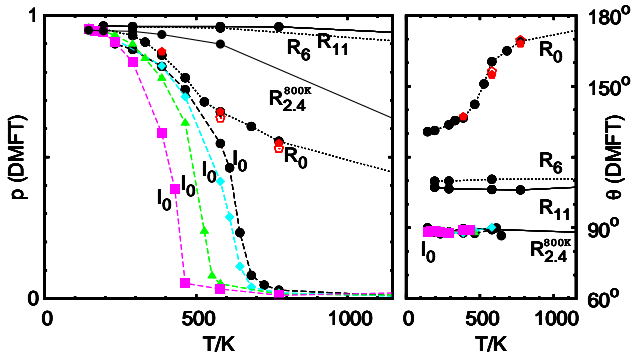}}
 \caption{\label{pol} (Color online)
  Orbital polarization $p$ (left) and (right) occupied state  $|\theta\rangle$ 
  ($|-\theta\rangle$) for sites 1 and 3 (2 and 4)
  as a function of temperature. 
  Solid line:  300~K (R$_{11}$) and 800~K (R$_{2.4}^{\rm 800K}$) structures. 
  Dots: orthorhombic structures with half (R$_6$) or no (R$_0$) Jahn-Teller distortion.
  Pentagons: 2 (full) and 4 ( empty) sites
  CDMFT. Dashes: ideal cubic structure (I$_0$).
  Circles: $U=5$~eV. Diamonds: $U=5.5$~eV. Triangles: $U=6$~eV. Squares: $U=7$~eV.
  Crystal-field splitting (meV): 840  (R$_{11}$), 495  (R$_6$), 219  (R$_0$),
  168 (R$_{2.4}^{\rm 800K}$), and 0 (I$_0$).} 
\end{figure}

For the 300~K structure (R$_{11}$) the band-widths are $W_{t_{2g}}\!\sim\!1.6$~eV and $W_{e_g}\!\sim\! 3.0$~eV. The $e_g$ states split by $\sim840$~meV, in good agreement with experimental estimates \cite{cef}. The lower crystal-field state at site 1 is $|1\rangle=0.574|3z^2-1\rangle+0.818|x^2-y^2\rangle$. 
We find an insulating solution in the full range $U=4-7$~eV (Fig.~\ref{dmftdos}). The Mott gap  $E_g$ is $\sim 0.6$~eV for $U=4$~eV, and increases almost linearly with increasing $U$.  For $U=5$~eV, suggested by recent estimates \cite{hero,hubbard1},
the Hubbard bands are at $\sim -1.5$ and 2~eV. In addition there is a broad feature around $5$~eV due to $e_g$ states with spin antiparallel to the randomly oriented t$_{2g}$ spins. These spectra are in line with experiments \cite{cef,opticalgaps,hubbard1,hubbard2}.
We find that even at 1150~K the system is fully  orbitally polarized ($p\sim1$). On sites 1 and 3, the occupied state is $|\theta\rangle\sim|106^o\rangle$, on sites 2 and 4 it is $|-\theta\rangle\sim|-106^o\rangle$ ($d$-type OO); $|\theta\rangle$ is close to the lower crystal-field state obtained from LDA (table \ref{hops}) and in excellent agreement with neutron diffraction experiments \cite{rodriguez98}.
We find that things hardly change when the JT distortion is halved (R$_{6}$ structure in Fig.~\ref{pol}).
Even for the average 800~K structure (R$_{2.4}^{\rm 800K}$) OO does not disappear: Although the Jahn-Teller distortion is strongly reduced to $\delta_{JT}=2.4\%$, the crystal-field splitting is  $\sim$168~meV and the orbital polarization at 1150~K is as large as $p\sim 0.65$, while $\theta$ is now close to $90^o$. For all these structures, orbital order is already determined by the distortions via the crystal-field splitting.

To find the temperature $T_{\rm KK}$ at which Kugel-Khomskii super-exchange drives orbital-order we consider the zero crystal-field limit, i.e.\ the ideal cubic structure, I$_0$. The $e_g$ band-width increases to $W_{e_g}\sim 3.7$~eV and for $U=5$~eV the system is a Mott insulator with a tiny gap only below $T\sim 650$~K. We find  $T_{\rm KK}\sim650$~K, very close to the metal-insulator transition (Fig.~\ref{pol}). To check how strongly $T_{\rm KK}$ changes when the gap opens, we increase $U$. For $U=5.5$~eV we find an insulating solution with a small gap of $\sim0.5$~eV and $T_{\rm KK}$ still close to $\sim650$~K. For $U=6$~eV, $E_g\sim0.9$~eV and $T_{\rm KK}\sim550$~K. Even with an unrealistically large $U=7$~eV, giving $E_g\sim1.8$~eV, $T_{\rm KK}$ is still as large as $\sim470$~K.
Thus, despite the small gap, $T_{\rm KK}$ decreases as $\sim1/U$, as expected for super-exchange. For a realistic $U\sim5$~eV, the calculated $T_{\rm KK}\sim650$~K is surprisingly close to the order-disorder transition temperature, $T_{OO}\sim 750~$K, though still much smaller than $T_{JT}\gtrsim1150$~K. The occupied state at site 1 is $|\theta\rangle\sim|90^o\rangle$ for all $U$.

Such a large $T_{\rm KK}$ is all the more surprising when compared with the value obtained for KCuF$_3$, $T_{\rm KK}\sim350$~K \cite{kcuf3}. 
For the ideal cubic structure the hopping matrix  (table~\ref{hops}) is 
$t^{i,i\pm{\bf z}}_{m,m'}\sim -t \delta_{m,m'}\delta_{m,3z^2-1}$,
$t^{i,i\pm{\bf x}}_{m,m}=t^{i,i\pm{\bf y}}_{m,m}\sim -t/4  (1+2\delta_{m,x^2-y^2})$, and for $m\neq m'$
 $t^{i,i\pm{\bf x}}_{m,m'}=-t^{i,i\pm{\bf y}}_{m,m'}\sim \sqrt{3}t/4$.
Since  the effective (after averaging over the
directions of ${\bf S}_{t_{2g}}$) hopping integral  in LaMnO$_3$,  $2t/3\sim 345$~meV is $\sim10\%$ smaller
than 
$\; t \sim 376$~meV in KCuF$_3$  \cite{kcuf3}, one may expect 
a slightly smaller $T_{\rm KK}$ in LaMnO$_3$, opposite to what we find.
Our result can, however, be understood in super-exchange theory.
The KK SE part of the Hamiltonian, obtained by second-order perturbation theory 
in $t$ from Eq.~(\ref{H}), may be written as
\begin{eqnarray} \label{SE}
\nonumber
H_{{SE}}^{i,i'}\!\!
&\sim& 
\frac{J_{SE}}{2}
\!\!\!\!\sum_{\langle ii'\rangle_{\bf x,y}}\!\!\!
    \left[3 T^x_iT^x_{i'}  
                    \mp \sqrt3 \left(T^z_iT^x_{i'}+T^x_iT^z_{i'}\right)\right]\\
&+&    \frac{J_{SE}}{2}
\!\!\!\!\sum_{\langle ii'\rangle_{\bf x,y}}\!\!\!
    T^z_iT^z_{i'}+  2J_{SE} \sum_{\langle ii'\rangle_{\bf z}} T^z_iT^z_{i'},  
\end{eqnarray}
where $\langle i,i'\rangle_{\bf x,y}$ and $\langle i,i'\rangle_{\bf z}$
indicate near neighboring sites
along {\bf x}, {\bf y}, or {\bf z}; $- (+)$ refers to the {\bf x} ({\bf y}) direction, $T^x_i$ and $T^z_i$ are pseudospin operators \cite{kugel},  with
$T^z|3z^2-1\rangle=1/2|3z^2-1\rangle$, $T^z|x^2-y^2\rangle=-1/2|x^2-y^2\rangle$.
The superexchange coupling is $J_{SE}=({\bar{t}^2}/{U}) (w/2)$,
where $\bar{t}$ is the effective hopping integral. In the large $U$ limit (negligible $J/U$ and $h/U$), 
$w\sim1 \!+\! 4\langle S_i^z\rangle 
\langle S_{i^\prime}^z\rangle \!+\! (1\!-\! 4 \langle S_i^z \rangle\langle S_{i^\prime}^z\rangle) u^{i,i^\prime}_{\Uparrow,\Downarrow} / u^{i,i^\prime}_{\Uparrow,\Uparrow}$, where $S_i^z$ are the $e_g$ spin operators.
In LaMnO$_3$ the $e_g$ spins align with the 
randomly oriented $t_{2g}$ spins, thus $\bar{t}=2t/3$,  $w\sim 2$,
and  $J_{SE}\sim 2 ({2t}/{3})^2 / U$.
For $d$-type order, the classical ground-state  is $|\theta\rangle\sim| 90^o\rangle$, in agreement with our DMFT
results. In KCuF$_3$, with configuration  $t_{2g}^6$e$_g^3$, 
the Hund's rule coupling between $e_g$  and $t_{2g}$ plays no role, i.e. $\langle S_i^z\rangle=0$.
The hopping integral $\bar{t}=t$ is  indeed slightly larger than in LaMnO$_3$, but $w \sim 1$, a reduction of $50\%$. 
Consequently, $J_{SE}$ is reduced by $\sim 0.6$ in KCuF$_3$.
For finite $J/U$ and $h/U$, $w$ is a more complicated function, but the conclusions stay the same.
We verified solving (\ref{H}) with LDA+DMFT  that also for LaMnO$_3$ T$_{\rm KK}$ drops drastically  
if $u^{i,i^\prime}_{\sigma,-\sigma}=0$ and $h=0$.

\begin{table}
 \begin{tabular}{c@{\hspace{4ex}}rrrr@{\hspace{6ex}}rrrr}
  $lmn$ &
  $t_{\pi,\pi\phantom{0} }^{i,i'}$ &
  $t_{\pi,0  \phantom{pi}}^{i,i'}$ & 
  $t_{0,\pi}^{i,i'}$ & 
  $t_{0,0  }^{i,i'}$ & 
  $t_{\pi,\pi\phantom{0} }^{i,i'}$ & 
  $t_{\pi,0  \phantom{pi}}^{i,i'}$ & 
  $t_{0,\pi}^{i,i'}$ & 
  $t_{0,0}^{i,i'}$\\\hline
         &\multicolumn{4}{c}{R$_{11}$} &\multicolumn{4}{c}{R$_{2.4}^{\rm 800K}$}\\
  $000$  &     0 &  409 &  409 &  305  &     0 &   84 &   84 &   -2\\
  $001$  &    -8 &  -47 &  -47 & -445  &    -2 &  -13 &  -13 & -439\\
  $010$  &  -322 &  233 &  174 & -129  &  -328 &  196 &  190 & -105\\  
  $100$  &  -322 & -174 & -236 & -129  &  -328 & -190 & -196 & -105\\[0.5ex]  
         &\multicolumn{4}{c}{R$_{0}$}  &\multicolumn{4}{c}{I$_{0}$}\\
  $000$  &     0 &    5 &    5 &  218  &     0 &    0 &    0 &    0\\
  $001$  &    -1 &   -2 &   -2 & -433  &   -10 &    0 &    0 & -518\\
  $010$  &  -333 &  206 &  207 & -121  &  -391 &  220 &  220 & -137\\  
  $100$  &  -333 & -207 & -206 & -121  &  -391 & -220 & -220 & -137
 \end{tabular}
 \caption[]{\label{hops} 
Hopping integrals $t_{m,m'}^{i,i'}$/meV from a site $i$ of type 1 to a neighboring site $i'$ of type 2 in direction $l{\bf x}+n{\bf y}+m {\bf z}$ for structures R$_{11}$, R$_{2.4}^{\rm 800K}$, R$_{0}$ and I$_0$. The states $m,m'$ are $|\pi\rangle=|x^2-y^2\rangle$ and $|0\rangle=|3z^2-1\rangle$. The crystal-field states are the eigenvectors of the on-site  matrix ($l=m=n=0$).}
\end{table}

It remains to evaluate the effect of the orthorhombic distortion on the  transition. For this we perform calculations for the system R$_{0}$ with no Jahn-Teller distortion, but keeping the tetragonal and GdFeO$_3$-type distortion of the 300~K structure. This structure is metallic for $U=4$~eV; for $U=5$~eV it has a gap of $\sim 0.5$~eV. We find a large polarization already at 1150~K ($p\sim 0.45$). Such polarization is due to the crystal-field splitting of about $219$~meV, with lower crystal-field states at site 1 given by $|1\rangle\sim | x^2-y^2\rangle$. Surprisingly the most occupied state $|\theta\rangle$ is close to
$|1\rangle$ ($\theta\sim 180)$  only at high temperature ($\sim 1000$~K). The orthorhombic crystal-field thus competes with super-exchange, analogous to an external field with a component perpendicular to an easy axis. On cooling the occupied orbitals rotate to $|\theta\rangle\sim |132^o\rangle$ (see Fig.~\ref{pol}). This effect of super-exchange, occurs around a characteristic temperature $T^{\rm R}_{\rm KK}\sim 550$~K; still surprisingly large, but reduced compared to $T_{\rm KK}$ for the ideal cubic system I$_0$ and much smaller than the experimental $T_{JT}\gtrsim 1150$~K.
Short-range correlations could reduce T$^{\rm R}_{\rm KK}$ or modify $\theta$. To estimate this effect we perform CDMFT calculations; our results (Fig.~\ref{pol}) remain basically unchanged.
Thus, electron-phonon coupling is necessary to explain both the transition temperature and the correct occupied orbital $|\theta\rangle\sim|108^o\rangle$.

In conclusion, we find that $T^{\rm R}_{\rm KK}$ in orthorhombic LaMnO$_3$ is $\sim550$~K. We have shown that two elements are crucial: the super-exchange mechanism, which yields a transition temperature as high as 650~K, and the tetragonal plus GdFeO$_3$-type distortion, which, due to the reduced hopping integrals and the competing orthorhombic crystal-field, reduces $T_{\rm KK}$ to 550~K. Experimentally, an order-to-disorder transition occurs around $T_{OO}\sim 750$~K, but a local
Jahn-Teller distortion persists in the disordered phase up to $T_{JT}\gtrsim 1150$~K. The Kugel-Khomskii mechanism alone cannot account for the presence of such  Jahn-Teller distortions above 550~K ($T^{\rm R}_{\rm KK}\ll T_{JT}$). It also cannot justify the neutron scattering estimate $\theta=108^o$. Thus electron-phonon coupling is a crucial ingredient, both for making the Jahn-Teller distortions energetically favorable at such high temperatures and in determining the occupied orbital.

We acknowledge 
discussions with  D.I.~Khomskii and A.I.~Lichtenstein, 
and grant JIFF22 on Jugene.


\begin{thebibliography}{99}
\bibitem{details}  Structure R$_{11}$, parameters $1/k_BT=5$ eV, $U=5$~eV, $J=0.75$~eV. We find $\sim$ t$_{2g}^3$e$_g^1$ and Hubbard bands in line with $S_{t_{2g}}=3/2$; the $e_g$ orbital polarization, $p$ and the gap are close to the results for Hamiltonian (\ref{H}).
\bibitem{fm} 
With $e_g$ spins alinged to the $t_{2g}$ spins,
the SE coupling (see discussion after Eq.~\ref{SE}) is
$J_{SE}\sim 2 \bar{t}^2 (1/(U-3J) + (u_{\Uparrow,\Downarrow}/u_{\Uparrow,\Uparrow}) (1/(U+2h-2J) -1/(U+2h))$,
i.e. depends weakly on the $u^{ii'}_{\sigma-\sigma}$ terms in  (\ref{H});
thus we neglect them.
\end{thebibliography}
\end{document}